# Fully Automatic Wound Segmentation with Deep Convolutional Neural Networks

**Chuanbo Wang[1], DM Anisuzzaman[1], Victor Williamson[1], Mrinal Kanti Dhar[1], Behrouz Rostami[2], Jeffrey Niezgoda[3], Sandeep Gopalakrishnan[4*] and Zeyun Yu[1, 5*]**

[1] Big Data Analytics and Visualization Laboratory, Department of Computer Science, University of Wisconsin-Milwaukee, Milwaukee, WI, USA;
[2]Department of Electrical Engineering, University of Wisconsin-Milwaukee, Milwaukee, WI, USA;
[3]Advancing the Zenith of Healthcare (AZH) Wound and Vascular Center, Milwaukee, WI, USA;
[4]College of Nursing, University of Wisconsin Milwaukee, Milwaukee, WI, USA;
[5]Department of Biomedical Engineering, University of Wisconsin-Milwaukee, Milwaukee, WI, USA.
[*]Corresponding authors:
Zeyun Yu, Associate Professor and Director of Big Data Analytics and Visualization Laboratory, Department of Computer Science, University of Wisconsin-Milwaukee, Milwaukee, WI, USA. Email: yuz@uwm.edu
Sandeep Gopalakrishnan, Assistant Professor, College of Nursing, University of Wisconsin Milwaukee, Milwaukee, WI, 53211, USA. Email: sandeep@uwm.edu

## ABSTRACT

Acute and chronic wounds have varying etiologies and are an economic burden to healthcare systems around the world. The advanced wound care market is expected to exceed $22 billion by 2024. Wound care professionals rely heavily on images and image documentation for proper diagnosis and treatment. Unfortunately lack of expertise can lead to improper diagnosis of wound etiology and inaccurate wound management and documentation. Fully automatic segmentation of wound areas in natural images is an important part of the diagnosis and care protocol since it is crucial to measure the area of the wound and provide quantitative parameters in the treatment. Various deep learning models have gained success in image analysis including semantic segmentation. Particularly, MobileNetV2 stands out among others due to its lightweight architecture and uncompromised performance. This manuscript proposes a novel convolutional framework based on MobileNetV2 and connected component labelling to segment wound regions from natural images. We build an annotated wound image dataset consisting of 1,109 foot ulcer images from 889 patients to train and test the deep learning models. We demonstrate the effectiveness and mobility of our method by conducting comprehensive experiments and analyses on various segmentation neural networks. The full implementation is available at https://github.com/Pele324/ChronicWoundSeg.

# Introduction

Acute and chronic nonhealing wounds represent a heavy burden to healthcare systems, affecting millions of patients around the world [1]. In the United States, medicare cost projections for all wounds are estimated to be between $28.1B and $96.8B [34]. Unlike acute wounds, chronic wounds fail to predictably progress through the phases of healing in an orderly and timely fashion, thus require hospitalization and additional treatment adding billions in cost for health services annually [2]. The shortage of well-trained wound care clinicians in primary and rural healthcare settings decreases the access and quality of care to millions of Americans. Accurate measurement of the wound area is critical to the evaluation and management of chronic wounds to monitor the wound healing trajectory and to determine future interventions. However, manual measurement is time-consuming and often inaccurate which can cause a negative impact on patients. Wound segmentation from images is a popular solution to these problems that not only automates the measurement of the wound area but also allows efficient data entry into the electronic medical record to enhance patient care.

Related studies on wound segmentation can be roughly categorized into two groups: traditional computer vision methods and deep learning methods. Studies in the first group focus on combining computer vision techniques and traditional machine learning approaches. These studies apply manually-designed feature extraction to build a dataset that is later used to support machine learning algorithms. Song et al. described 49 features that are extracted from a wound image using K-means clustering, edge detection, thresholding, and region growing in both grayscale and RGB [3]. These features are filtered and prepared into a feature vector that is used to train a Multi-Layer Perceptron (MLP) and a Radial Basis Function (RBF) neural network to identify the region of a chronic wound. Ahmad et al. proposed generating a Red-Yellow-Black-White (RYKW) probability map of an input image with a modified hue-saturation-value (HSV) model [4]. This map then guides the segmentation process using either optimal thresholding or region growing. Hettiarachchi et al. demonstrated an energy minimizing discrete dynamic contour algorithm applied on the saturation plane of the image in its HSV color model [5]. The wound area is then calculated from a flood fill inside the enclosed contour. Hani et al. proposed applying an Independent Component Analysis (ICA) algorithm to the pre-processed RGB images to generate hemoglobin-based images, which are used as input of K-means clustering to segment the granulation tissue from the wound images [6]. These segmented areas are utilized as an assessment of the early stages of ulcer healing by detecting the growth of granulation tissue on ulcer bed. Wantanajittikul et al. proposed a similar system to segment the burn wound from images [7]. Cr-Transformation and Luv-Transformation are applied to the input images to remove the background and highlight the wound region. The transformed images are segmented with a pixel-wise Fuzzy C-mean Clustering (FCM) algorithm. These methods suffer from at least one of the following limitations: 1) As in many computer vision systems, the hand-crafted features are affected by skin pigmentation, illumination, and image resolution, 2) They depend on manually tuned parameters and empirically handcrafted features which does not guarantee an optimal result. Additionally, they are not immune to severe pathologies and rare cases, which are very impractical from a clinical perspective, and 3) The performance is evaluated on a small biased dataset.

Since the successes AlexNet [28] achieved in the 2012 Imagenet large scale visual recognition challenge [29], the application of deep learning [27] in the domain of computer vision sparked interests in semantic segmentation [30] using deep convolutional neural networks (CNN) [8]. Typically, traditional machine learning and computer vision methods make decisions based on feature extraction. To segment the region of interest, one must guess a set of important features and then handcraft sophisticated algorithms that capture these features [9]. However, a CNN integrates feature extraction and decision making. The convolutional kernels of CNN extract the features and their importance is determined during the training of the network. In a typical CNN architecture, the input are processed by a sequence of convolutional layers and the output is gernerated by a fully connected layer that requires fixed-sized input. One successful variant of CNN is fully convolutional neural networks (FCN) [10]. A FCN is composed of convolutional layers without a fully connected layer as the output layer. This allows arbitrary input sizes and prevents the loss of spatial information caused by the fully connected layers in CNNs. Several FCN-based methods have been proposed to solve the wound segmentation problem. For example, Wanget al. estimated the wound area by segmenting wounds [11] with the vanilla FCN architecture [10]. With time-series data consisting of the estimated wound areas and corresponding images, wound healing progress is predicted using a Gaussian process regression function model. However, the mean Dice accuracy of the segmentation is only evaluated to be 64.2%. Goyal et al. proposed to employ the FCN-16 architecture on the wound images in a pixel-wise manner that each pixel of an image is predicted to which class it belongs [12]. The segmentation result is simply derived from the pixels classified as a wound. By testing different FCN architectures they are able to achieve a Dice coefficient of 79.4% on their dataset. However, the network's segmentation accuracy is limited in distinguishing small wounds and wounds with irregular borders as the tendency is to draw smooth contours. , Liu et al. proposed a new FCN architecture that replaces the decoder of the vanilla FCN with a skip-layer concatenation upsampled with bilinear interpolation [13]. A pixel-wise softmax layer is appended to the end of the network to produce a probability map, which is post-processed to be the final segmentation. A dice accuracy of 91.6% is achieved on their dataset with 950 images taken under an uncontrolled lighting environment with a complex background. However, images in their dataset are semi-automatically annotated using a watershed algorithm. This means that the deep learning model is learning how the watershed algorithm labels wounds as opposed to human specialists.

To better explore the capacity of deep learning on the wound segmentation problem, we propose an efficient and accurate framework to automatically segment wound regions. The segmentation network of this framework is built above MobileNetsV2 [14]. This network is light-weight and computationally efficient since significantly fewer parameters are used during the training process.

Our contributions can be summarized as follows:

1. We build a large dataset of wound images with segmentation annotations done by wound specialists. This is by far the largest dataset focused on wound segmentation (to the best of our knowledge).
2. We propose a fully automatic wound segmentation framework based on MobileNetsV2 that balances computational efficiency and accuracy.
3. Our proposed framework shows high efficiency and accuracy in wound image segmentation.

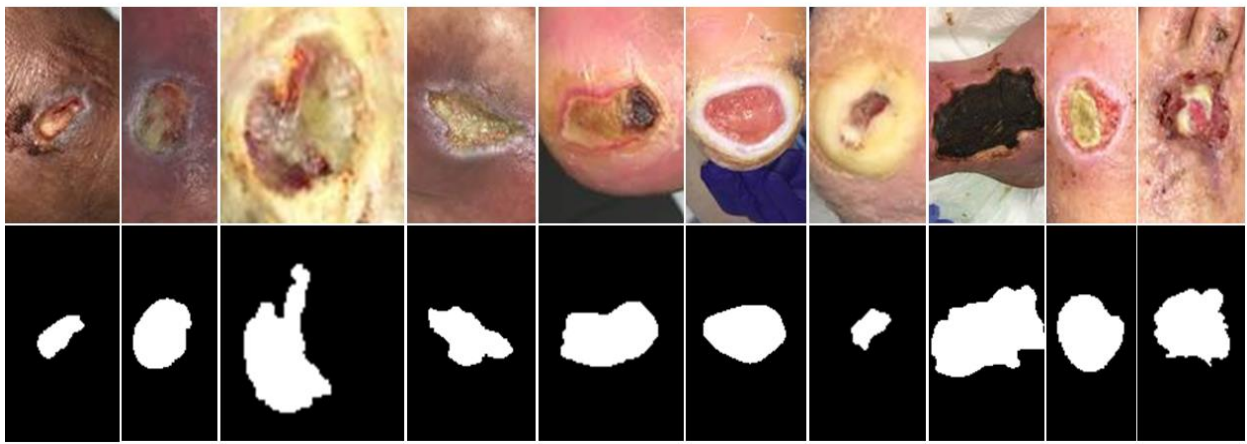

**Figure 1.** An illustration of images in our dataset. The first row contains the raw images collected. The second row consists of segmentation mask annotations we create with the AZH wound and vascular center.

## Dataset

**Dataset Construction**

There is currently no public dataset large enough for training deep-learning-based models for wound segmentation. To explore the effectiveness of wound segmentation using deep learning models, we collaborated with the Advancing the Zenith of Healthcare (AZH) Wound and Vascular Center, Milwaukee, WI. Our chronic wound dataset was collected over 2 years at the center and includes 1,109 foot ulcer images taken from 889 patients during multiple clinical visits. The raw images were taken by Canon SX 620 HS digital camera and iPad Pro under uncontrolled illumination conditions, with various backgrounds. Fig. 1 shows some sample images in our dataset.

The raw images collected are of various sizes and cannot be fed into our deep learning model directly since our model requires fixed-size input images. To unify the size of images in our dataset, we first localize the wound by placing bounding boxes around the wound using an object localization model we trained de novo, YOLOv3 [15]. Our localization dataset contains 1,010 images, which are also collected from the AZH Wound and Vascular Center. We augmented the images and built a training set containing 3645 images and a testing set containing 405 images. For training our model we have used LabelImg [24] to manually label all the data (both for training and testing). The YOLO format has been used for image labelling. The model has been trained with a batch size of 8 for 273 epochs. With an intersection over union (IoU) rate of 0.5 and non-maximum suppression of 1.00, we get the mean Average Precision (mAP) value of 0.939. In the next step, image patches are cropped based on the bounding boxes result from the localization model. We unify the image size (224 pixels by 224 pixels) by applying zero-padding to these images, which are regarded in our dataset data points. We confirm that the data collected was de-identified and in accordance to relevant guidelines and regulations and the patient's informed consent is waived by the institutional review board of University of Wisconsin-Milwaukee.

**Data Annotation**

During training, a deep learning model is learning the annotations of the training dataset. Thus, the quality of annotations is essential. Automatic annotation generated with computer vision algorithms is not ideal when deep learning models are trained to learn how human experts recognize the wound region. In our dataset, the images were manually annotated with segmentation masks that were further reviewed and verified by wound care specialists from the collaborating wound clinic. Initially only foot ulcer images were annotated and included in the dataset as these wounds tend to be smaller than other types of chronic wounds, which makes it easier and less time-consuming to manually annotate the pixel-wise segmentation masks. In the future we plan to create larger image libraries to include all types of chronic wounds, such as venous leg ulcers, pressure ulcers, and surgery wounds as well as non-wound reference images. The AZH Wound and Vascular Center, Milwaukee, WI, had consented to make our dataset publicly available.

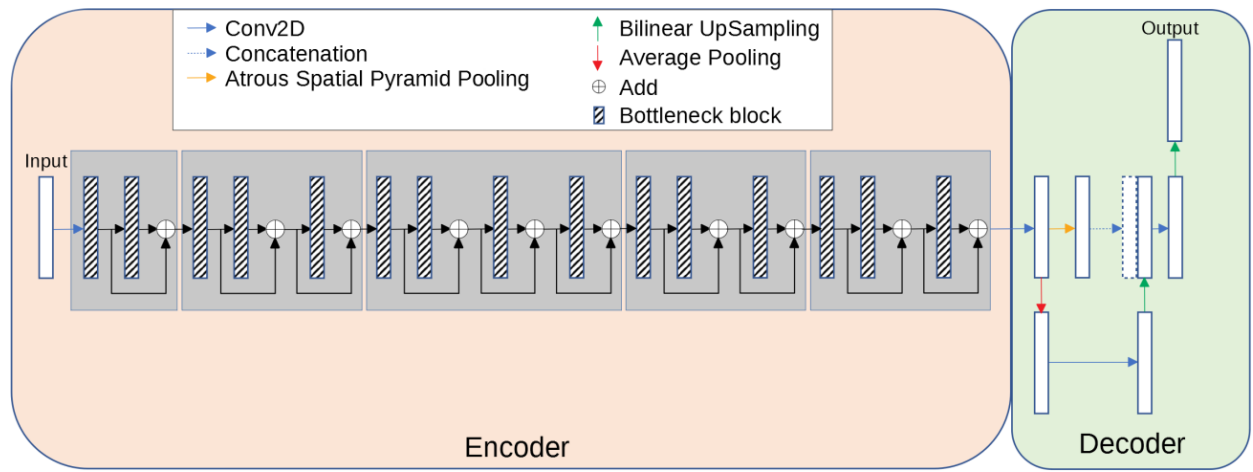

**Figure 2.** The encoder-decoder architecture of MobilenetV2.

## Methods

In this section we describe our method with the architecture of the deep learning model for wound segmentation. The transfer learning used during the training of our model and the post-processing methods including hole filling and removal of small noises are also described. We confirm that the research is approved by the institutional review board of University of Wisconsin-Milwaukee.

**Model Architecture Overview**

A convolutional neural network (CNN), MobileNetV2 [14], is adopted to segment the wound from the images. Compared with conventional CNNs, this network substitutes the fundamental convolutional layers with depth-wise separable convolutional layers [31] where each layer can be separated into a depth-wise convolution layer and a point-wise convolution layer. A depth-wise convolution performs lightweight filtering by applying a convolutional filter per input channel. A point-wise convolution is a $1 \times 1$ convolution responsible for building new features through linear combinations of the input channels. This substitution reduces the computational cost compared to traditional convolution layers by almost a factor of $k^2$ where k is the convolutional kernel size. Thus, depth-wise separable convolutions are much more computationally efficient than conventional convolutions suitable for mobile or embedded applications where computing resource is limited. For example, the mobility of MobileNetV2 could benefit medical

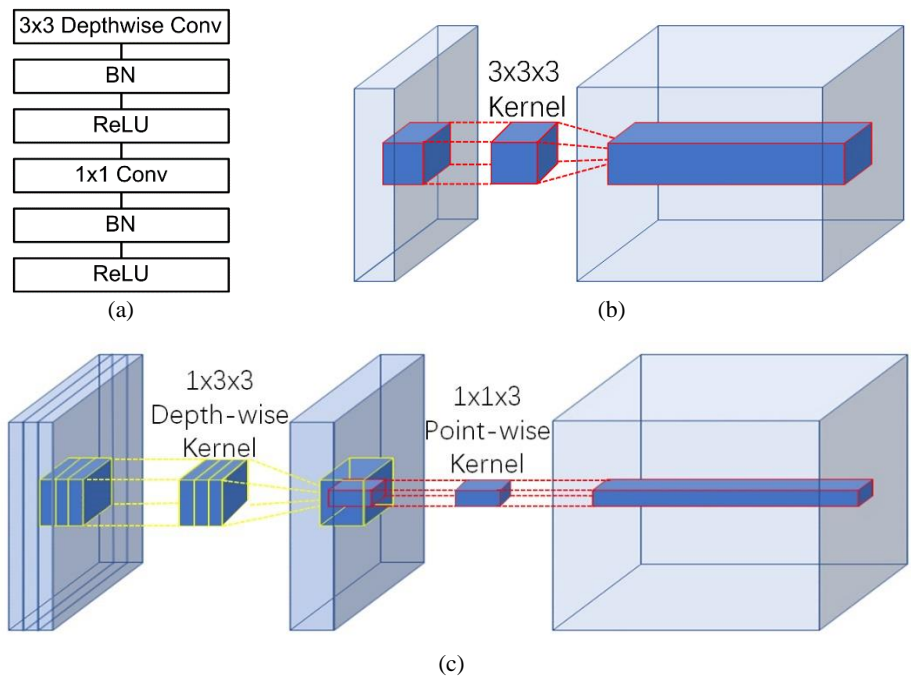

**Figure 3 (a).** A depth-separable convolution block. The block contains a $3 \times 3$ depth-wise convolutional layer and a $1 \times 1$ point-wise convolution layer. Each convolutional layer is followed by batch normalization and Relu6 activation. **(b)** An example of a convolution layer with a $3 \times 3 \times 3$ kernel. **(c)** An example of a depth-wise separable convolution layer equivalent to (b).

professionals and patients by allowing instant wound segmentation and wound area measurement immediately after the photo is taken using mobile devices like smartphones and tablets. An example of a depth-wise separable convolution layer is shown in Figure 3(c), compared to a traditional convolutional layer shown in Figure 3(b).

The model has an encoder-decoder architecture as shown in Figure 2. The encoder is built by repeatedly applying the depth-separable convolution block (marked with diagonal lines in Figure 2). Each block, illustrated in Figure 3(a), consists of six layers: a $3 \times 3$ depth-wise convolutional layer followed by batch normalization and Relu activation [32], and a $1 \times 1$ point-wise convolution layer followed again by batch normalization and Relu. To be more specific, Relu6 [33] was used as the activation function. In the decoder, shown in Figure 2, the encoded features are captured in multiscale with a spatial pyramid pooling block, and then concatenated with higher-level features generated from a pooling layer and a bilinear up-sampling layer. After the concatenation, we apply a few $3 \times 3$ convolutions to refine the features followed by another simple bilinear up-sampling by a factor of 4 to generate the final output. A batch normalization layer is inserted into each bottleneck block and a dropout layer is inserted right before the output layer. In MobileNetV2, a width multiplier α is introduced to deal with various dimensions of input images. we let α = 1 thus the input image size is set to 224 pixels × 224 pixels in our model.

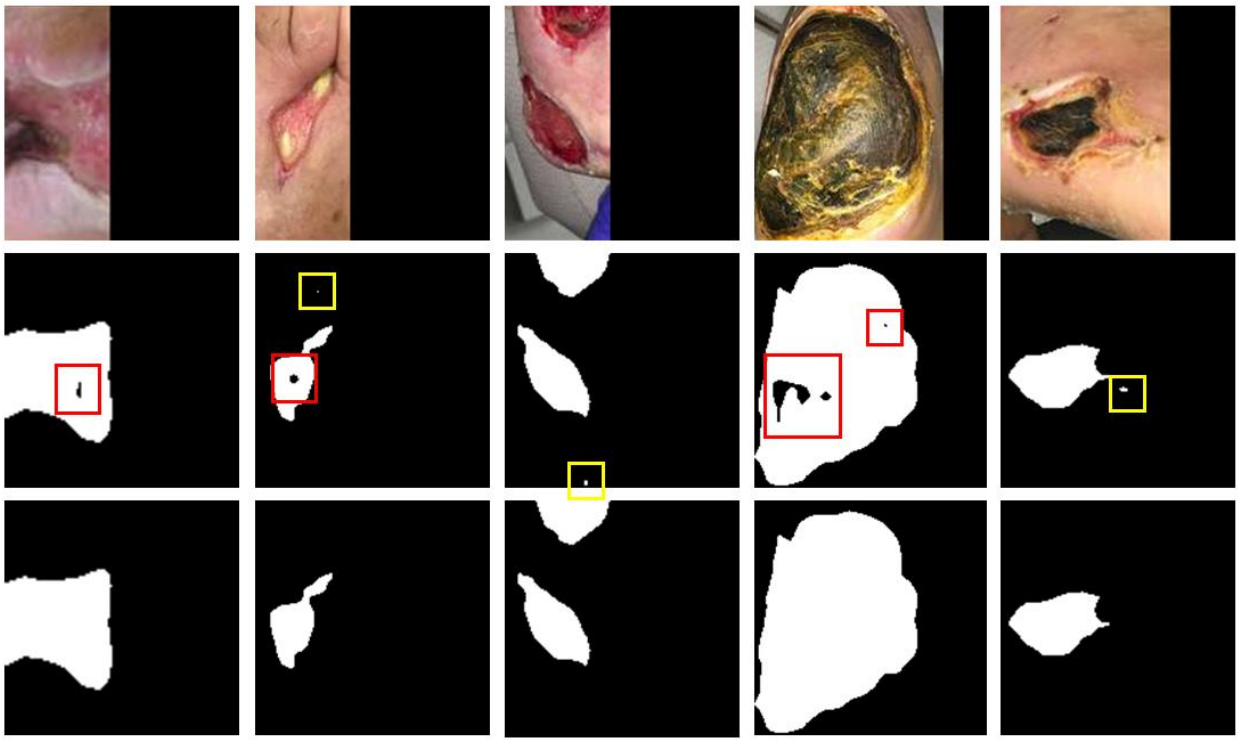

**Figure 4.** An illustration of the segmentation result and the post processing method. The first row illustrates images in the testing dataset. The second row shows the segmentation results predicted by our model without any post processing. The holes are marked with red boxes and the noises are marked with yellow boxes. The third row shows the final segmentation masks generated by the post processing method.

**Transfer Learning**

To make the training more efficient, we used transfer learning for our deep learning model. Instead of randomly initializing the weights in our model, the MobileNetV2 model, pre-trained on the Pascal VOC segmentation dataset [16] was loaded before training. Transfer learning with the pre-trained model is beneficial to the training process in the sense that the weights converge faster and better.

**Post-processing**

The raw segmentation masks predicted by our model are grayscale images with pixel intensities that range from 0 to 255. In the post processing step, binary segmentation masks are first generated from thresholding with a fixed threshold of 127, which is half the max intensity. The binary masks are further processed by hole filling and removal of small regions to generate the final segmentation masks as shown in Figure 4. We notice that abnormal tissue like fibrinous tissue within chronic wounds could be identified as non-wound and cause holes in the segmented wound regions. Such holes are detected by finding small connected components in the segmentation results and filled to improve the true positive rate using connected component labelling (CCL) [26]. The small false-positive noises are removed in the same way. The images in the dataset are cropped from the raw image for each wound. So, we simply remove noises in the segmentation results by removing the connected component small enough based on adaptive thresholds. To be more specific, a connected region is removed when the number of pixels within the region is less than a threshold, which is adaptively calculated based on the total number of pixels segmented as wound pixels in the image.

# Results

We describe the evaluation metrics and compare the segmentation performance of our method with several popular and state-of-the-art methods. Our deep learning model is trained with data augmentation and preprocessing. Extensive experiments were conducted to investigate the effectiveness of our network. FCN-VGG-16 is a popular network architecture for wound image segmentation [12] [21]. Thus, we trained this network on our dataset as the baseline model. For fairness of comparison, we used the same training strategies and data augmentation strategies throughout the experiments.

**Evaluation metrics**

To evaluate the segmentation performance, Precision, Recall, and the Dice coefficient are adopted as the evaluation metrics[25]:

*Precision*: Precision shows the accuracy of segmentation. More specifically, Precision measures the percentage of correctly segmented pixels in the segmentation and is computed by:

$$\text{Precision} = \frac{True\ positives}{True\ positives + False\ positives}$$

*Recall*: Recall also shows the accuracy of segmentation. More specifically, it measures the percentage of correctly segmented pixels in the ground truth and is computed by:

$$\text{Recall} = \frac{True\ positives}{True\ positives + False\ negtives}$$

*Dice coefficient (Dice)*: Dice shows the similarity between the segmentation and the ground truth. Dice is also called F1 score as a measurement balancing Precision and Recall. More specifically, Dice is computed by the harmonic mean of Precision and Recall:

$$\text{Dice} = \frac{2\ \times\ True\ positives}{2\ \times\ True\ positives + False\ negtives + False\ positives}$$

**Experiment setup**

The deep learning model in the presented work was implemented in python with Keras [17] and Tensorflow [18] backend. To speed up the training, the models were trained on a 64-bit Ubuntu PC with an 8-core 3.4GHz CPU and a single NVIDIA RTX 2080Ti GPU. For updating the parameters in the network, we employed the Adam optimization algorithm [19], which has been popularized in the field of stochastic optimization due to its fast convergence compared to other optimization functions. Binary cross entropy was used as the loss function and we also monitored Precision, Recall, and the Dice score as the evaluation matrices. The initial learning rate was set to 0.0001 and each minibatch contained only 2 images for balancing the training accuracy and efficiency. The convolutional kernels of our network were initialized with HE initialization [20] to speed up the training process and the training time of a single epoch took about 77 seconds. We used early stopping to terminate the training so that the best result was saved when there was no improvement for more than 100 epochs in terms of Dice score. Eventually, our deep learning model was trained for around 1000 epochs before overfitting.

**Table 1.** The precision, recall, and dice score evaluated using various models on our dataset.

| Model | VGG16 | SegNet | U-Net | Mask-RCNN | MobileNetV2 | MobileNetV2+CCL |
|---|---|---|---|---|---|---|
| Precision | 83.91% | 83.66% | 89.04% | **94.30%** | 90.86% | 91.01% |
| Recall | 78.35% | 86.49% | **91.29%** | 86.40% | 89.76% | 89.97% |
| Dice | 81.03% | 85.05% | 90.15% | 90.20% | 90.30% | **90.47%** |

To evaluate the performance of the proposed method, we compared the segmentation results achieved by our methods with those by FCN-VGG-16 [12][21], SegNet [11], and Mask-RCNN [35][36]. We also added 2D U-Net [22] to the comparison due to its outstanding segmentation performance on biomedical images with a relatively small training dataset. The segmentation results predicted by our model are demonstrated in Figure 4 along with the illustration of our post processing method. Quantitative results evaluated with the different networks are presented in Table 1 where bold numbers indicate the best results among all the models.

Table 2. The precision, recall, and dice score evaluated using various models on the Medetec dataset.

| Model | VGG16 | SegNet | U-Net | Mask-RCNN | MobileNetV2 | MobileNetV2+CCL |
|---|---|---|---|---|---|---|
| Precision | 77.84% | 72.03% | 86.84% | **98.40%** | 93.69% | 93.84% |
| Recall | 80.69% | 73.87% | 81.33% | 88.60% | 94.06% | **94.27%** |
| Dice | 79.24% | 72.94% | 84.01% | 93.20% | 93.88% | **94.05%** |

**Comparing our method to the others**

In the performance measures, the Recall of our model was evaluated to be the second highest among all models, at 89.97%. This was 1.32% behind the highest Recall, 91.29%, which was achieved by U-Net. Our model also achieved the second highest Precision of 91.01%. Overall, the results show that our model achieves the highest accuracy with a mean Dice score of 90.47%. the VGG16 was shown to have the worst performance among all the other CNN architectures. Mask-RCNN achieved the highest Precision of 94.30%, which indicates that the segmentation predicted by Mask-RCNN contains the highest percentage of true positive pixels. However, the Recall is only evaluated to 86.40%, meaning that more false negative pixels are undetected compared to U-Net and MobileNetV2. Our accuracy was slightly higher than U-Net and Mask-RCNN, and significantly higher than SegNet and VGG16.

**Comparison within the Medetec Dataset**

Apart from our dataset, we also conducted experiments on the Medetec Wound Dataset [23] and compared the segmentation performance of these methods. The results are shown in Table 2. We annotated the dataset in the same way that our dataset was annotated and trained the networks with the same experimental setup. The highest Dice score is evaluated to 94.05% using MobileNetV2+CCL. The performance evaluation agrees with the conclusion drawn from our dataset where our method outperforms the others regardless of which chronic wound segmentation dataset is used, thereby demonstrating that our model is robust and unbiased.

Table 3. Comparison of total numbers of trainable parameters.

| Model Name | FCN-VGG16 | SegNet | U-Net | Mask-RCNN | MobileNetV2 |
|---|---|---|---|---|---|
| Number of parameters | 134,264,641 | 902,561 | 4,834,839 | 63,621,918 | 2,141,505 |

## Discussion

Comparing our method to VGG16, the Dice score is boosted from 81.03% to 90.47% tested on our dataset. Based on the appearance of chronic wounds, we know that wound segmentation is complicated by various shapes, colors, and the presence of different types of tissue. The patient images captured in clinic settings also suffer from various lighting conditions and perspectives. In MobileNetV2, the deeper architecture has more convolutional layers than VGG16, which makes MobileNetV2 more capable to understand and solve these variables. MobileNetV2 utilizes residual blocks with skip connections instead of the sequential convolution layers in VGG16 to build a deeper network. These skip connections bridging the beginning and the end of a convolutional block allows the network to access earlier activations that weren't modified in the convolutional block and enhance the capacity of the network.

Another comparison between U-Net and SegNet indicates that the former model is significantly better in terms of mean Dice score. Similar to the previous comparison, U-Net also introduces skip connections between convolutional layers to replace the pooling indices operation in the architecture of SegNet. These skip connections concatenate the output of the transposed convolution layers with the feature maps from the encoder at the same level. Thus, the expansion section which consists of a large number of feature channels allows the network to propagate localization combined with contextual information from the contraction section to higher resolution layers. Intuitively, in the expansion section or "decoder" of the U-Net architecture, the segmentation results are reconstructed with the structural features that are learned in the contraction section or the "decoder". This allows the U-Net to make predictions at more precise locations. These comparisons have illustrated the effectiveness of skip connections for improving the accuracy of wound segmentation.

Besides the performance, our method is also efficient and lightweight. As shown in Table 3, the total number of trainable parameters in the adopted MobileNetV2 was only a fraction of the numbers in U-Net, VGG16, and Mask-RCNN. Thus, the network took less time during training and could be applied to mobile devices with less memory and limited computational power. Alternatively, higher-resolution input images could be fed into MobileNetV2 with less memory size and computational power comparing to the other models.

## Conclusions

We attempted to solve the automated segmentation problem of chronic foot ulcers in a dataset we built on our own using deep learning. We conducted comprehensive experiments and analyses on SegNet, VGG16, U-Net, Mask-RCNN, and our model based on MobileNetV2 and CCL to evaluate the performance of chronic wound segmentation. In the comparison of various neural networks, our method has demonstrated its effectiveness and mobility in the field of image segmentation due to its fully convolutional architecture consisting of depth-wise separable convolutional layers. We demonstrated the robustness of our model by testing it on the foot ulcer images in the publicly available Medetec Wound Dataset where our model still achieves the highest Dice score. In the future, we plan to improve our work by a novel multi-stream neural network architecture that extracts the shape features separately from the pixel-wise convolution in our deep learning model. Also, we will include more data in the dataset to improve the robustness and prediction accuracy of our method.

## References

[1] Frykberg, Robert G., and Jaminelli Banks. "Challenges in the treatment of chronic wounds. *Advances in wound care*. **4**, no. 9, 560-582 (2015).

[2] Branski, Ludwik K., Gerd G. Gauglitz, David N. Herndon, and Marc G. Jeschke. A review of gene and stem cell therapy in cutaneous wound healing. *Burns*. **35**, no. 2, 171-180 (2009).

[3] Song, Bo, and Ahmet Sacan. Automated wound identification system based on image segmentation and artificial neural networks. *In 2012 IEEE International Conference on Bioinformatics and Biomedicine*. 1-4 (2012).

[4] Fauzi, Mohammad Faizal Ahmad et al. Computerized segmentation and measurement of chronic wound images. *Computers in biology and medicine*. **60**, 74-85 (2015).

[5] Hettiarachchi, N. D. J., R. B. H. Mahindaratne, G. D. C. Mendis, H. T. Nanayakkara, and Nuwan D. Nanayakkara. Mobile based wound measurement. *In 2013 IEEE Point-of-Care Healthcare Technologies (PHT)*. 298-301 (2013).

[6] Hani, Ahmad Fadzil M., Leena Arshad, Aamir Saeed Malik, Adawiyah Jamil, and Felix Yap Boon Bin. Haemoglobin distribution in ulcers for healing assessment. *In 2012 4th International Conference on Intelligent and Advanced Systems (ICIAS2012)*. **1**, 362-367 (2012).

[7] Wantanajittikul, Kittichai, Sansanee Auephanwiriyakul, Nipon Theera-Umpon, and Taweethong Koanantakool. Automatic segmentation and degree identification in burn colour images. *In The 4th 2011 Biomedical Engineering International Conference*. 169-173 (2012).

[8] LeCun, Yann, Léon Bottou, Yoshua Bengio, and Patrick Haffner. Gradient-based learning applied to document recognition. *Proceedings of the IEEE.* **86**, no. 11, 2278-2324 (1998).

[9] Wang, Chuanbo, Ye Guo, Wei Chen, and Zeyun Yu. Fully automatic intervertebral disc segmentation using multimodal 3d u-net. *In 2019 IEEE 43rd Annual Computer Software and Applications Conference (COMPSAC).* **1**, 730-739 (2019).

[10] Long, Jonathan, Evan Shelhamer, and Trevor Darrell. Fully convolutional networks for semantic segmentation. *In Proceedings of the IEEE conference on computer vision and pattern recognition.* 3431-3440 (2015).

[11] Wang, Changhan et al. A unified framework for automatic wound segmentation and analysis with deep convolutional neural networks. *In 2015 37th annual international conference of the IEEE engineering in medicine and biology society (EMBC).* 2415-2418 (2015).

[12] Goyal, Manu, Moi Hoon Yap, Neil D. Reeves, Satyan Rajbhandari, and Jennifer Spragg. Fully convolutional networks for diabetic foot ulcer segmentation. *In 2017 IEEE international conference on systems, man, and cybernetics (SMC).* 618-623 (2017).

[13] Liu, Xiaohui et al. A framework of wound segmentation based on deep convolutional networks. *In 2017 10th International Congress on Image and Signal Processing, BioMedical Engineering and Informatics (CISP-BMEI).* 1-7 (2017).

[14] Sandler, Mark, Andrew Howard, Menglong Zhu, Andrey Zhmoginov, and Liang-Chieh Chen. Mobilenetv2: Inverted residuals and linear bottlenecks. *In Proceedings of the IEEE conference on computer vision and pattern recognition.* 4510-4520 (2018).

[15] Redmon, Joseph, and Ali Farhadi. Yolov3: An incremental improvement. Preprint at arXiv:1804.02767 (2018).

[16] Everingham et al. The pascal visual object classes challenge: A retrospective. *International journal of computer vision.* **111**, no. 1, 98-136 (2015).

[17] F. Chollet et al, Keras, chollet2015keras, https://keras.io.

[18] S. S. Girija, Tensorflow: Large-scale machine learning on heterogeneous distributed systems. (2016).

[19] Kingma, Diederik P., and Jimmy Ba. Adam: A method for stochastic optimization. Preprint at arXiv:1412.6980 (2014).

[20] He, Kaiming, Xiangyu Zhang, Shaoqing Ren, and Jian Sun. Delving deep into rectifiers: Surpassing human-level performance on imagenet classification. *In Proceedings of the IEEE international conference on computer vision*. 1026-1034 (2015).

[21] Li, Fangzhao, Changjian Wang, Xiaohui Liu, Yuxing Peng, and Shiyao Jin. A composite model of wound segmentation based on traditional methods and deep neural networks. *Computational intelligence and neuroscience.* (2018).

[22] Ronneberger, Olaf, Philipp Fischer, and Thomas Brox. U-net: Convolutional networks for biomedical image segmentation. *In International Conference on Medical image computing and computer-assisted intervention*. 234-241 (2015).

[23] Thomas, Stephen. Stock Pictures of Wounds. *Medetec Wound Database* http://www.medetec.co.uk/files/medetec-image-databases.html (2020).

[24] Tzutalin. LabelImg. *Git code* https://github.com/tzutalin/labelImg (2015).

[25] Zou, Kelly H. et al, Statistical validation of image segmentation quality based on a spatial overlap *index1: scientific reports. Academic radiology.* **11**, no. 2, 178-189 (2004).

[26] Pearce, David J. An improved algorithm for finding the strongly connected components of a directed graph. Victoria University, Wellington, NZ, Tech. Rep (2005).

[27] Litjens, Geert et al. A survey on deep learning in medical image analysis. *Medical image analysis.* **42,** 60-88 (2017).

[28] Krizhevsky, Alex, Ilya Sutskever, and Geoffrey E. Hinton. Imagenet classification with deep convolutional neural networks. *In Advances in neural information processing systems*. 1097-1105 (2012).

[29] Russakovsky, Olga et al. Imagenet large scale visual recognition challenge. *International journal of computer vision.* **115**, no. 3, 211-252 (2015).

[30] Garcia-Garcia, Alberto, Sergio Orts-Escolano, Sergiu Oprea, Victor Villena-Martinez, and Jose Garcia-Rodriguez. A review on deep learning techniques applied to semantic segmentation. arXiv preprint arXiv:1704.06857 (2017).

[31] Chollet, François. Xception: Deep learning with depthwise separable convolutions. *In Proceedings of the IEEE conference on computer vision and pattern recognition*. 1251-1258 (2017).

[32] Nair, Vinod, and Geoffrey E. Hinton. Rectified linear units improve restricted boltzmann machines. *In Proceedings of the 27th international conference on machine learning (ICML-10)*. 807-814 (2010).

[33] Krizhevsky, Alex, and Geoff Hinton. Convolutional deep belief networks on cifar-10. *Unpublished manuscript*. **40**, 1-9 (2010).

[34] Sen, Chandan K. Human wounds and its burden: an updated compendium of estimates. *Advances in Wound Care*. **8**, 39-48 (2019).

[35] He, Kaiming, et al. Mask r-cnn. *Proceedings of the IEEE international conference on computer vision*. (2017).

[36] Abdulla, Waleed. Mask R-CNN for object detection and instance segmentation on Keras and TensorFlow. *Git code* https://github.com/matterport/Mask_RCNN (2017)